\documentclass[onecolumn,showpacs,aps,amssymb,floatfix,prd,amsmath,preprintnumbers]{revtex4}
\usepackage{capt-of}
\usepackage{graphicx}  
\usepackage{dcolumn}   
\RequirePackage{graphicx}
\RequirePackage{float}
\RequirePackage{amsmath}
\RequirePackage{amssymb}
\RequirePackage{mathtools}
\usepackage[utf8]{inputenc}

\begin{document}

\title{\bf Cosmic acceleration and energy conditions in $\Lambda$CDM symmetric teleparallel  $f(Q)$ gravity}
\author{ M. Koussour$^{1}$\footnote{Email: pr.mouhssine@gmail.com},
  S. H. Shekh$^{2}$\footnote{Email: da\_salim@rediff.com},
M. Bennai$^{3}$\footnote{Email: mdbennai@yahoo.fr}}\

\affiliation{$^{1}$ Quantum Physics and Magnetism Team, LPMC,Faculty of Science Ben M'sik, Casablanca Hassan II University, Morocco.}
\affiliation{$^{2}$ Department of Mathematics, S.P.M. Science and Gilani Arts and Commerce College, Ghatanji, Yavatmal, Maharashtra-445301, India.}
\affiliation{$^{3}$ Lab of High Energy Physics, Modeling and Simulations,Faculty of Science, University Mohammed V-Agdal, Rabat, Morocco.}

\begin{abstract}
\textbf{Abstract:}In this paper, we study a spatially homogeneous and isotropic FLRW cosmological model in framework of the symmetric teleparallel $f(Q)$ gravity proposed by Jimenez et al. (Physical Review D 98.4 (2018): 044048), where the non-metricity $Q$ is responsible for the gravitational interaction. For the whole analysis we have considered a special form of $f\left(Q\right)$ gravity function i.e. $f\left( Q\right) =\alpha Q^{n}+\beta $ (where $\alpha $, $\beta $, and $n$ are the dynamical model constant parameters) in the direction of $t\left( z\right) =\frac{kt_{0}}{b}f\left( z\right)$, where, $f\left( z\right) =W\left[ \frac{b}{k}e^{\frac{b-\ln \left(1+z\right) }{k}}\right] $ and $W$ denotes the Lambert function. We have discussed two cases of the function: linear $f(Q)$ model towards $ n=1$ and quadratic for $n=2$. In both the model we had verified the validity of the models with the help of energy conditions along with some physical parameters like equation of state parameter, jerk parameter, statefinder parameters are also discussed in details and compared it with the observational data.
\end{abstract}

\pacs{04.50+h}

\maketitle

\textbf{Keywords}: FRW model; $f(Q)$ gravity; energy conditions; cosmology.

\section{Introduction}

The phenomenon of the current accelerated expansion of the Universe is considered one of the confusing cosmic mysteries currently in the scientific area, which has created a great controversy among researchers today. Because the subject which is related to a group of astronomical and cosmological observations, all of which point to a result, namely that our current Universe is entering a phase of acceleration [1 - 4]. This great controversy was a contradiction between the observation data and the theory that explains the large scale, i.e. the theory of General Relativity (GR). To have an appropriate theoretical explanation of this phenomenon, various alternative theories have recently been proposed by many researchers. A modified gravity theories (MGT) are among the most common, which have recently emerged as alternatives to GR, to explain the current problem of cosmic acceleration, all acting with the principle of modifying the Einstein Hilbert (EH) action by replacing the Ricci scalar $R$ by a more general function, which may be of the Ricci scalar or any other function with matter-geometry coupling. Here, are some of the most used MGT: $f\left(R\right) $ gravity ($R$ is Ricci scalar) [5, 6], $f\left(T\right) $ gravity ($T$ is torsion scalar) [7 - 12], $f\left(G\right) $ gravity ($G$ is Gauss-Bonnet scalar) [13 - 15], $f\left( R,T\right) $ gravity ($R$ is Ricci scalar and $T$ is the trace of energy-momentum tensor) [16 - 20], $f\left(T,B\right) $ gravity ($T$ be the torsion scalar and $B$ is the boundary term), $f(R,G)$ modified theory of gravity[21, 22].\\
Recently, a new modified theory of gravity has attracted the interest of researchers called symmetric teleparallel (ST) gravity or $f(Q)$ gravity proposed by Jimenez et al. [23], where the non-metricity term $Q$ is responsible to the gravitational interaction. The essential difference between ST and GR is the role contributed by the affine connection, $\Gamma _{\mu \nu }^{\alpha }$ instead of the physical manifold. Also, $f\left( Q\right) $ gravity is equivalent to GR in flat space. It is important to keep in mind that same to the $f\left( T\right) $ gravity, $f\left( Q\right) $ gravity also nature in second-order field equations, while gravitational field equations of $f\left( R\right) $ gravity are of the fourth-order. Motivated by the ease of solving field equations, much work has been done in this context, we mention: Khyllep et al. [24] studied the power-law form of $f(Q)$ gravity i.e. $f\left( Q\right) =Q+\alpha Q^{n}$ to obtain the late-time acceleration of the Universe while Mandal et al. [25] discuss the validity of $f(Q)$ gravity using energy conditions for several cosmological models and found that these models are compatible with the accelerated expansion of our Universe, the role of bulk viscosity to study the accelerated expansion of the Universe in the framework of modified $f(Q)$ gravity was also studied by Solanki et al. [26], and many other works that were born in $f\left( Q\right) $ gravity, see, for example [27 - 30].\\
The energy conditions are a set of conditions that describe matter in the Universe and are used in many approaches to understanding the evolution of the Universe. In this work, the role of energy conditions is to verify the acceleration of the expansion of the Universe. Such conditions can be derived from the well-known Raychaudhury equations, whose forms are $\frac{d\theta }{d\tau }=-\frac{1}{3}\theta ^{2}-\sigma _{\mu \nu }\sigma^{\mu \nu }+\omega _{\mu \nu }\omega ^{\mu \nu }-R_{\mu \nu }u^{\mu }u^{\nu}$ and $\frac{d\theta }{d\tau }=-\frac{1}{2}\theta ^{2}-\sigma _{\mu \nu }\sigma^{\mu \nu }+\omega _{\mu \nu }\omega ^{\mu \nu }-R_{\mu \nu }n^{\mu }n^{\nu}$, where $\theta $ is the expansion factor, $n^{\mu }$ is the null vector, and $\sigma ^{\mu \nu }$ and $\omega _{\mu \nu }$ are, respectively, the shear and the rotation associated with the vector field $u^{\mu }$. The attractive gravity satisfy the following energy conditions
\begin{itemize}
\item Weak energy conditions (WEC) if $\rho \geq 0,\rho +p\geq 0$,
\item Null energy condition (NEC) if $p+\rho \geq 0$,
\item Dominant energy conditions (DEC) if $\rho \geq 0,\left\vert p\right\vert \leq \rho $,
\item Strong energy condition (SEC) if $\rho +3p \geq 0$.
\end{itemize}
In the analysis of the energy conditions show that a violation of NEC leads to a violation of the remaining energy conditions, which represents the depletion of energy density with expansion of the Universe, moreover the violation of SEC represents the acceleration in the expansion of the Universe [21, 25].\\
By motivating and inspiring with the work mentioned in the above references in this work, we construct a cosmological model that describes the evolution of the Universe on a large scale, we will work within the framework of the isotropic and spatially homogeneous Friedmann-Lema\^{\i}tre-Robertson-Walker (FLRW) Universe with the perfect fluid matter by considering the  model $f(Q)=\alpha Q^{n}+\beta $ [31] towards a linear and quadratic case respectively for $\left( n=1\right) $ and  $\left( n=2\right) $ and also to verify the validity of the constructed model with the help of energy conditions in both cases.\\
The present paper is organized as follows: In Section II, we formulate the gravitational field equations of $f\left( Q\right) $ gravity for a flat FLRW Universe. The cosmographic solution of the field equations are obtained in Section III. The dynamical properties along with energy conditions for both linear and quadratic cases  of this model are examined in the Sections IV while  jerk and statefinder parameters are discussed the in Sections, V and VI. Finally, we give our concluding remarks in the last Section VII.

\section{Basic formalism of the $f(Q)$ gravity}

In $f(Q)$ gravity, the action is given by [31]
\begin{equation}
S=\int \left[ -\frac{1}{2}f(Q)+L_{m}\right] d^{4}x\sqrt{-g},  \label{eqn1}
\end{equation}
where $f(Q)$ is an arbitrary function of the non-metricity $Q$, $g$ is the determinant of the metric tensor $g_{\mu \nu }$ and $L_{m}$ is the matter Lagrangian density. \\
The variation of the action $\left( S\right) $ in Eq. (1) with respect to the metric, we get the gravitational field equation given by
\begin{equation}
\frac{2}{\sqrt{-g}}\nabla _{\gamma }\left( \sqrt{-g}f_{Q}P^{\gamma }{}_{\mu
\nu }\right) +\frac{1}{2}fg_{\mu \nu }+f_{Q}\left( P_{\mu \gamma i}Q_{\nu
}{}^{\gamma i}-2Q_{\gamma i\mu }P^{\gamma i}{}_{\nu }\right) =T_{\mu \nu },
\label{eqn7}
\end{equation}
where $f_{Q}=\frac{df}{dQ}$. In addition, we can also take the variation of (1) with respect to the connection, which gives
\begin{equation}
\nabla _{\mu }\nabla _{\gamma }\left( \sqrt{-g}f_{Q}P^{\gamma }{}_{\mu \nu}\right) =0.  \label{eqn8}
\end{equation}
In this paper, we will study the isotropic and homogeneous Friedmann-Lema\^{\i}tre-Robertson-Walker (FLRW) Universe, with line element written in the form
\begin{equation}
ds^{2}=-dt^{2}+a^{2}(t)\left[ dr^{2}+r^{2}\left( d\theta ^{2}+\sin^{2}\theta d\phi ^{2}\right) \right] ,  \label{eqn9}
\end{equation}
where $a(t)$ is the scale factor of the Universe, $\left( t,r,\theta ,\phi\right) $ are the comoving coordinates.
The trace of the non-metricity tensor for the previous line element is found as
\begin{equation}
Q=6H^{2}.  \label{eqn10}
\end{equation}
Now, we take matter as a perfect fluid, so the energy-momentum tensor is given by
\begin{equation}
T_{\mu \nu }=\left( \rho +p\right) u_{\mu }u_{\nu }+pg_{\mu \nu },
\end{equation}
where $u_{\mu }$ is the four-velocity of the fluid satisfying the condition $u_{\mu }u^{\mu }=-1$, $p$ represents the isotropic pressure and $\rho $ represents the energy density. The equation of motion (2) for the FLRW model (4) with the fluid of energy-momentum tensor (6), we find the modified Friedmann equations for $f(Q)$ gravity as
\begin{equation}
3H^{2}=\frac{1}{2f_{Q}}\left( \rho +\frac{f}{2}\right) , 
\end{equation}
\begin{equation}
\overset{\cdot }{H}+3H^{2}+\frac{\overset{.}{f_{Q}}}{f_{Q}}H=\frac{1}{2f_{Q}}\left( -p+\frac{f}{2}\right) ,  \label{eqn13}
\end{equation}
where $H$ is the Hubble parameter defined by $H=\frac{\overset{\cdot }{a}}{a} $ and the overhead dot designates the derivative with respect to cosmic time $t$. Also, the field  equations towards Friedmann model for the energy density and pressure as
\begin{equation}
\rho =-\frac{f}{2}+6H^{2}f_{Q}, 
\end{equation}
\begin{equation}
p=\frac{f}{2}-\left( \overset{.}{H}+3H^{2}+\frac{\overset{.}{f_{Q}}}{f_{Q}}H\right) \left( 2f_{Q}\right).
\end{equation}
To find the exact solutions of the above system of equations, an additional constrain must be added for the scale factor $\left( a\right) $ or otherwise the deceleration parameter $\left( q\right) $. We will discuss this additional constrain under cosmographic solutionsin the next section.

\section{Cosmographic solutions}

In order to obtain a cosmological model where the Universe shifts from the early deceleration phase to the current accelerated phase, we will assume that the scale factor follows the following Hybrid Expansion Law (HEL) [32]
\begin{equation}
a\left( t\right) =a_{0}\left( \frac{t}{t_{0}}\right) ^{k}e^{b\left( \frac{t}{t_{0}}-1\right) },
\end{equation}
where $k\geq 0$, $b\geq 0$ are constant model parameters and $a_{0}$, $t_{0}$ represent the present values of scale factor and the age of the Universe, respectively.\\
The main motivation for choosing the scale factor in the above form is that it produces a time-dependent deceleration parameter, as indicated by the observation data. Also, it is known in the literature that this law is a generalization of both the power-law cosmology and the exponential law cosmology. If we take $k=0$ HEL reduces to exponential law of the form $a=a_{0}.e^{b.\left( \frac{t}{t_{0}}-1\right) }$ and if taken $b=0$, this reduces to power-law $a=a_{0}\left( \frac{t}{t_{0}}\right) ^{k}$. For this particular form of scale factor, the Hubble parameter and the deceleration parameter both are  respectively obtain as $(k/t)+(b/t_{0})$ and $-1+(kt_{0}^{2})/(bt+kt_{0}) ^{2}$. As, several recent observational data have shown that $q>0$ which describes a decelerating Universe, and a $q<0) $ describes the accelerating expansion of the Universe, other observational data from Type Ia Supernovae (SN Ia) has shown that the current Universe is in the acceleration phase by confirming the range of deceleration parameter $-1\leq q<0$. It is clear that there is a transition phase from deceleration to acceleration at $t=\frac{t_{0}}{b}\left( -k\pm \sqrt{k}\right) $ with $0<k<1$. Because the negativity of the second term leads to a concept of negative time, which appears to be unphysical in the context of Big Bang cosmology, we close that the cosmic transition may have occurred at $t=\frac{t_{0}}{b}\left( \sqrt{k}-k\right) $. Also, we notice that $q=-1$ for $k=0$ (exponential law) and $q=\left( 1/k\right) -1$ for $b=0$ (power-law).\\
The relation between the redshift $(z)$ and the scale factor is given by
\begin{equation}
a\left( t\right) =\frac{a_{0}}{1+z},  \label{eqn19}
\end{equation}
where $a_{0}$\ is the present value of scale factor.\\
The age of Universe at redshift is given as
\begin{equation}
t\left( z\right) =\frac{kt_{0}}{b}f\left( z\right) ,  \label{eqn20}
\end{equation}
where, $f\left( z\right) =W\left[ \frac{b}{k}e^{\frac{b-\ln \left(1+z\right) }{k}}\right] $ and $W$ denotes the Lambert function also known as \textquotedblleft product logarithm\textquotedblright.\\
With the help of Eqs. (12) and (13), the Hubble's parameter in redshift reads as,
\begin{equation}
H(z)=\frac{H_{0} b}{k+b}\left[\frac{1}{f(z)}+1\right]
\end{equation}
where $H_{0}=\frac{k+b}{t_{0}}$ be the present value of Hubble's parameter.\\
Keep in mind Eq. (14), the deceleration parameter in redshift reads as,
\begin{equation}
q\left( z\right) =-1+\frac{1}{k}\left[ \frac{1}{f\left( z\right) +1}\right] ^{2} 
\end{equation}
The plot of deceleration parameter versus redshift with the appropriate choice of constants is shown in fig.1.
we can easily see that the performance of  deceleration parameter is  from positive to negative in fig. 1 (a) (left panel) in fig. 1(b) (right panel) with respect to cosmic time and redshift respectively which produces an early deceleration phase followed by a current acceleration phase with the current values of the deceleration parameter for $k=0.6$ and $b=0.4, 0.7$. The behavior of deceleration parameter is more consistent with recent observational data. Also, it is look like as cosmic structure formation and thermodynamic arguments for $k=0.6$ and $b=0.4, 0.7$. The result of deceleration parameter in the present model is resumbles with result obtained mentioned in [29]. Hence fix the value of constants as $k=0.6$ and $b=0.4, 0.7$ throughout the analysis of all dynamical physical parameters.

\begin{figure}[H]
\begin{minipage}{0.49\linewidth}
\centerline{\includegraphics[scale=1]{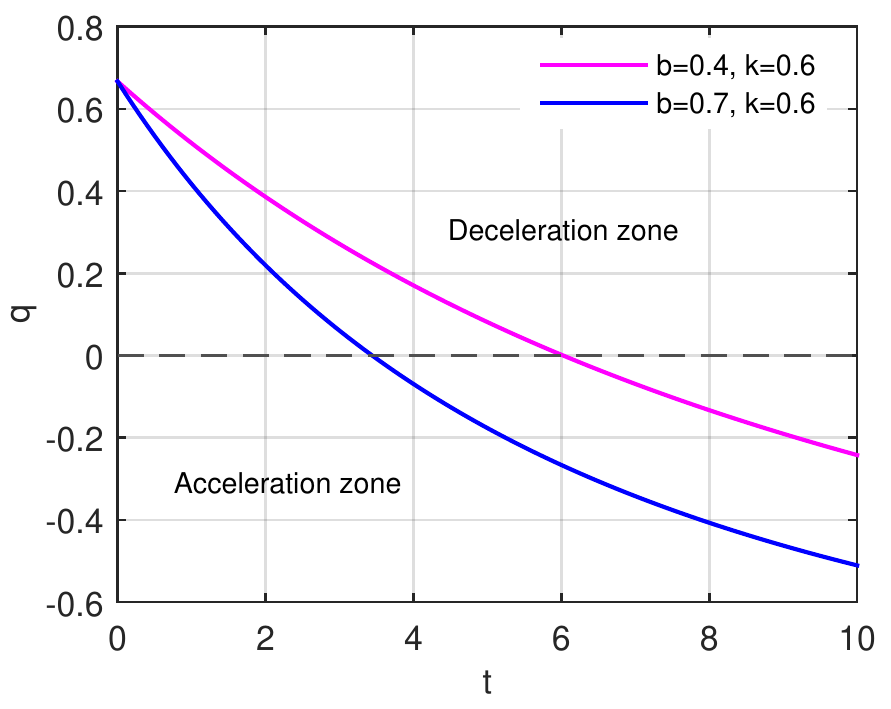}}
\centering{(a)}
\end{minipage}
\hfill
\begin{minipage}{0.49\linewidth}
\centerline{\includegraphics[scale=1]{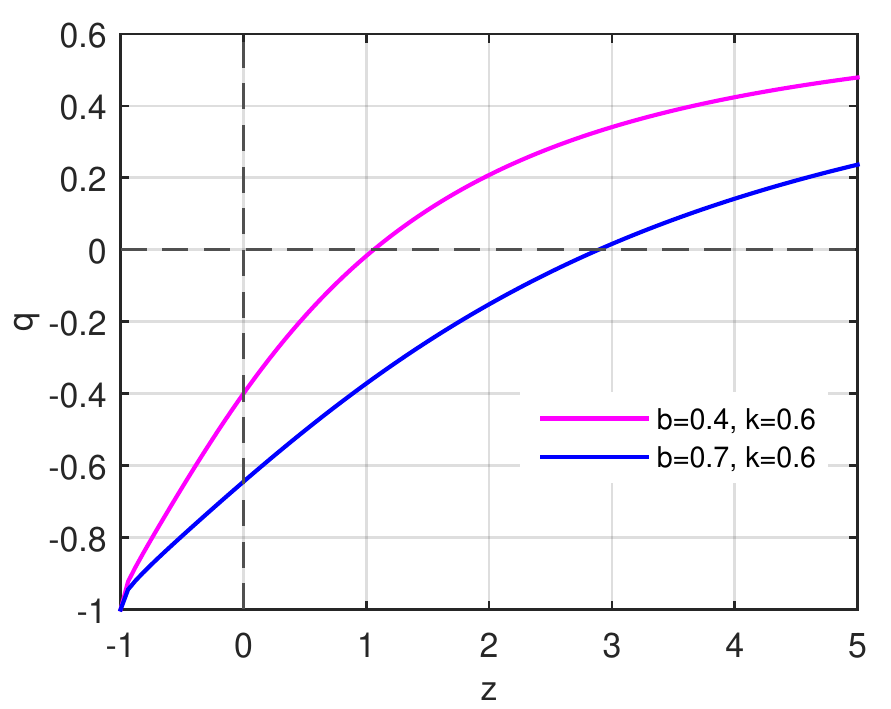}}
\centering{(b)}
\end{minipage}
\caption{Evolution of deceleration parameter in $f (Q)$ gravity model with the appropriate choice of constants $k=0.6$ and $b=0.4, 0.7$.}
\end{figure}
\section{Dynamical properties of the model}
In order to obtain equations for the energy density $\rho $\ of the Universe and isotropic pressure $p$ in terms of redshift function, we assume the function $f\left( Q\right) $ of the form: $f\left( Q\right) =\alpha Q^{n}+\beta $, where $\alpha $, $\beta $, and $n$ are constants.\\
From the Eq. (9) and (10) using (14),  we obtain the expression of $\rho(z)$, $p(z)$ and $\omega(z)$ of the Universe as
\begin{equation}
\rho (z)=\frac{\alpha 6^{n+1}\left( 2n+1\right) }{2}H^{2\left( n+1\right) }- \frac{\beta }{2}, 
\end{equation}
\begin{equation}
p(z)=\frac{\beta }{2}-\alpha 6^{n}\left( 2n+1\right) \left[ 3+2\frac{\overset{.} {H}}{H^{2}}\left( n+1\right) \right] H^{2\left( n+1\right) }, 
\end{equation}

\begin{equation}
\omega(z)=\frac{2 \left(\frac{\beta }{2}-\alpha  6^n (2 n+1) H^{2 n+1} (3 H+2 \dot{H} (n+1))\right)}{\alpha  6^{n+1} (2 n+1) H^{2 n+2}-\beta }
\end{equation}
In order to study the dynamic properties of the model and to find the behavior of physical parameters, we discuss two simple cases in following two subsections: The first is the linear form with $n=1$, and the second is the non-linear form with $n=2$ to which the model parameters specified and in the case where $\alpha =-1$ and $\beta =n=0$ our model corresponds to standerd GR.\\
\subsection{Case-I: Linear model} 
The linear model takes the form $f(Q)=\alpha Q+\beta $. Towards this model from equations (16) to (18), the expressions of physical parameters such as energy density, isotropic pressure and equation of state parameter along with energy conditions terms of redshift are obtained as,\\
Isotropic pressure as,
\begin{equation}
p=\frac{\beta }{2}- 18 \alpha \left[ 3+4\frac{\overset{.} {H}}{H^{2}}\right] H^{4}, 
\end{equation}
The graphical illustration of the isotropic pressure in the derived model with respect to redshift is depicted in fig.(2) by supposing the suitable choice of value of constants. From the figure it is detected that in derived model, an isotropic pressure has initially positive and with the evolution of the model it becomes negative, this negative pressure or repulsive force can be look upon as a source of the iflation which is the exact capable tool to explain a cosmic acceleration also it is interesting to take note that the same behavior of isotropic pressure resumbles with the fact that the current Universe is an accelerating expansion due to large negative pressure along with the analysis of Shekh et al. [35, 36].\\
 Energy density as,
\begin{equation}
\rho =54\alpha H^{4}- \frac{\beta }{2}, 
\end{equation}
\begin{figure}[H]
 \begin{minipage}{0.45\linewidth}
  \centerline{\includegraphics[scale=0.65]{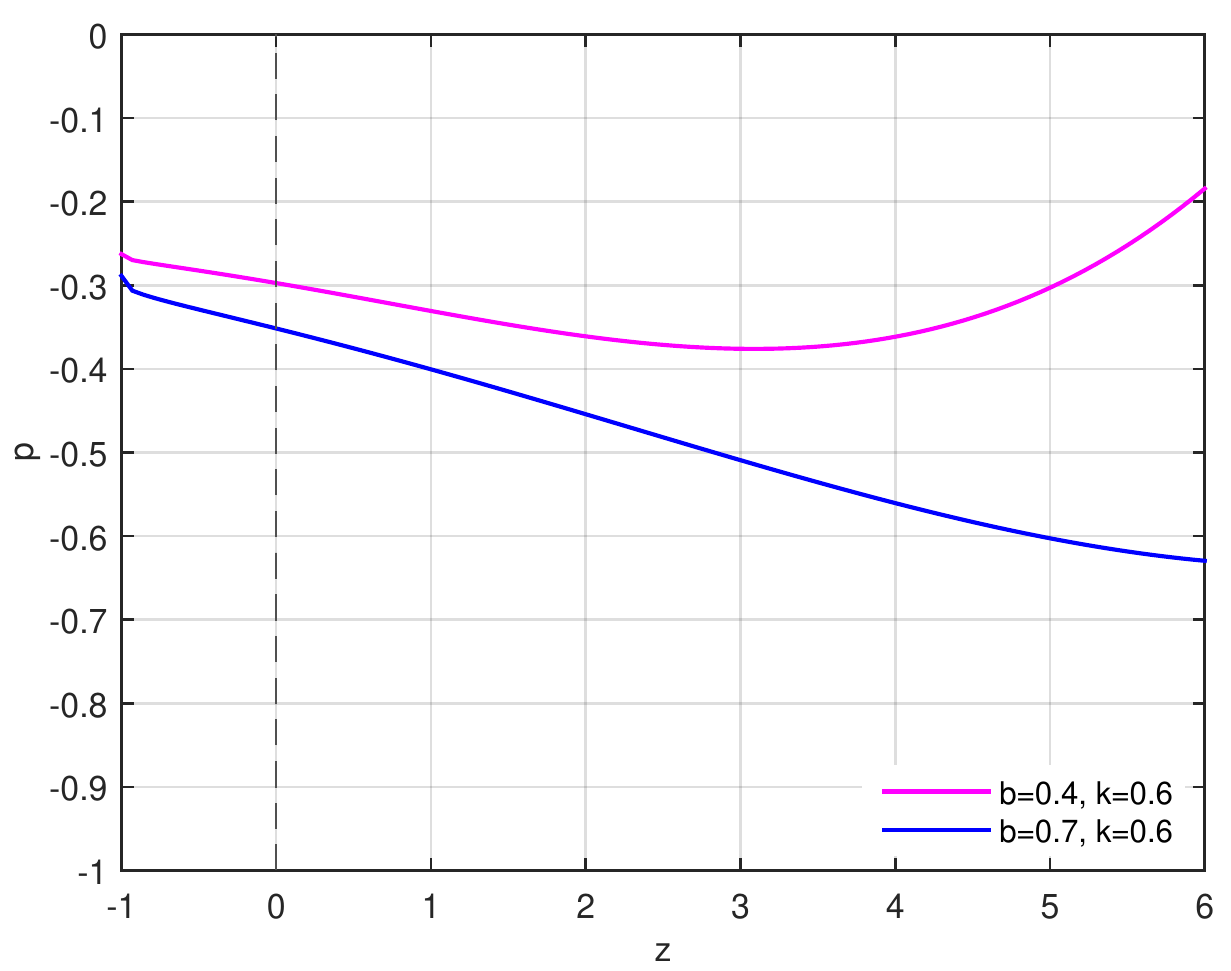}}
  \caption{Evolution of  isotropic pressure in linear form of $f (Q)$ gravity model versus redshift for the appropriate choice of constants $k=0.6$ and $b=0.4, 0.7$.}\label{fig3}
 \end{minipage}
\hfill
 \begin{minipage}{0.45\linewidth}
  \centerline{\includegraphics[scale=0.65]{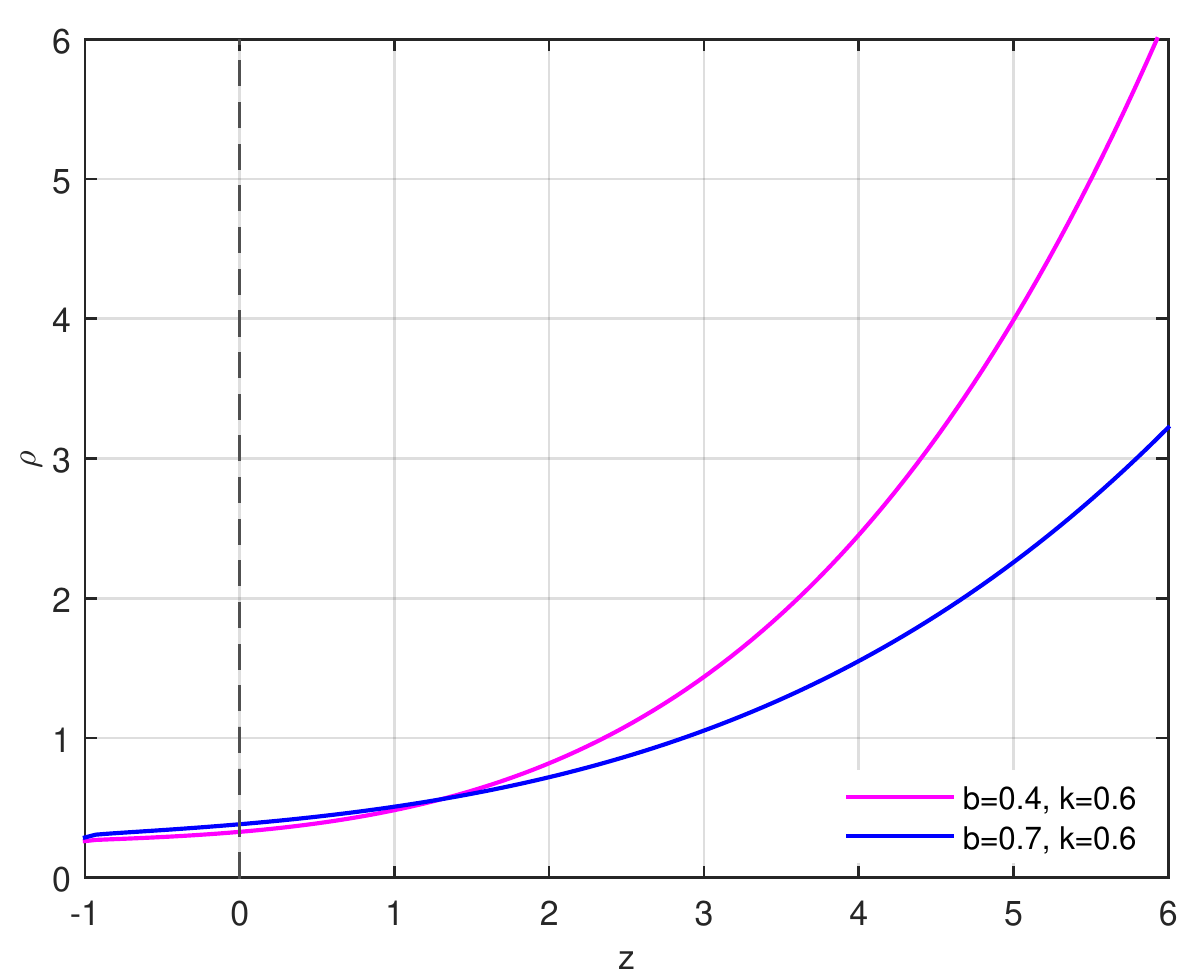}}
  \caption{Evolution of energy density in linear form of $f (Q)$ gravity model versus redshift for the appropriate choice of constants $k=0.6$ and $b=0.4, 0.7$.}\label{fig4}
 \end{minipage}
\end{figure}
The graphical illustration of the energy density of the derived model with respect to redshift is depicted in fig. (3) by supposing the suitable choice of values of constants. From the figure, it is observed that the energy density of linear $f(Q)$ gravity model provides an empty universe for large time as shown in fig. (3) which confirms the fact that volume of the space increases because of the density of matter decreases as the Universe expands. The behavior resumbles with the work of Shekh et al. [34].\\
Equation of state parameter as,
\begin{equation}
\omega=\frac{2 \left(\frac{\beta }{2}-18\alpha  H^{3} (3 H+4 \dot{H})\right)}{108 \alpha H^{4}-\beta }
\end{equation}
The graphical illustration of equation of state parameter of the derived model with respect to redshift is depicted in fig. (4) by supposing the suitable choice of values of constants.
\begin{figure}[H]
\centerline{\includegraphics[scale=0.9]{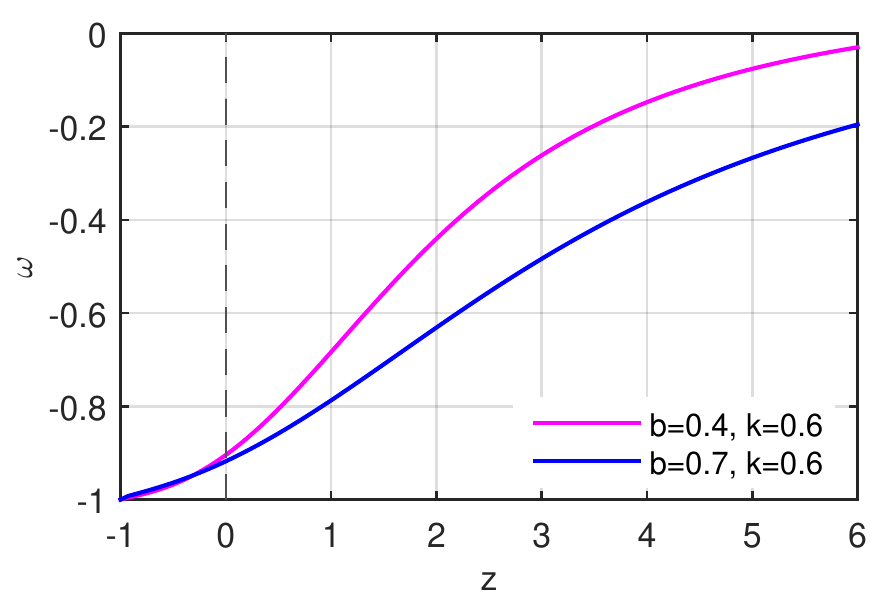}}
\caption{Evolution of equation of state parameter in linear form of $f (Q)$ gravity model versus redshift for the appropriate choice of constants $k=0.6$ and $b=0.4, 0.7$.}
\end{figure}
We can see that the equation of state parameter of a linear model is a function of redshift and converges to $\omega \rightarrow -1$ as time increases (see fig. 4). The equation of state parameter in linear model represents a quintessence region $\left( -1<\omega <-1/3\right) $ at present $z=0$.  Also, it can be seen that the  equation of state parameter is approaching the $\Lambda CDM$ model $\left( \omega =-1\right) $ in the future. This indicates that our model creates a very large accelerated phase for very weak redshift values. This behavior of equation of state parameter is resumbles with the recent observations.\\
Energy conditions as,
\begin{equation}
WEC\Longleftrightarrow 3 \alpha  H^2-\frac{\beta }{2}, 54 \alpha  H^4-18\alpha  H^4 \left(\frac{4 \dot{H}}{H^2}+3\right)
\end{equation}
\begin{equation}
NEC\Longleftrightarrow 54 \alpha  H^4-18\alpha  H^4 \left(\frac{4 \dot{H}}{H^2}+3\right)
\end{equation}
\begin{equation}
DEC\Longleftrightarrow 54\alpha  H^4-\left| \frac{\beta }{2}-18H^4 \left(\frac{4 \dot{H}}{H^2}+3\right) \alpha \right| -\frac{\beta }{2}
\end{equation}
\begin{equation}
SEC\Longleftrightarrow 54 \alpha  H^4+3 \left(\frac{\beta }{2}-18\alpha H^4 \left(\frac{4 \dot{H}}{H^2}+3\right)\right)-\frac{\beta }{2}
\end{equation}
Evolution of energy conditions  in linear form of linear $f (Q)$ gravity model versus redshift for the appropriate choice of constants is presented in fig. (5). The fig. describe the deeds of energy conditions such as WEC, NEC, DEC and SEC for the linear form of $f(Q)$ gravity with $\alpha >0$ and $\beta <0$. Also, from these two figures, we observed that WEC ($\rho \geq 0$) together with NEC ($\rho +p\geq 0$) and the DEC  ($\rho-\left\vert p\right\vert\ge 0$) are verified where as the  SEC ($\rho+3p\ge0$) violats. Therefore, the violation of SEC leads to the expansion of the Universe is accelerating. Also, it is seen that the SEC is violated at present $\left( z=0\right) $ and future $\left( z\rightarrow -1\right) $. Hence, as a result our model predicts an accelerating Universe.
\begin{figure}[H]
\centerline{\includegraphics[scale=0.9]{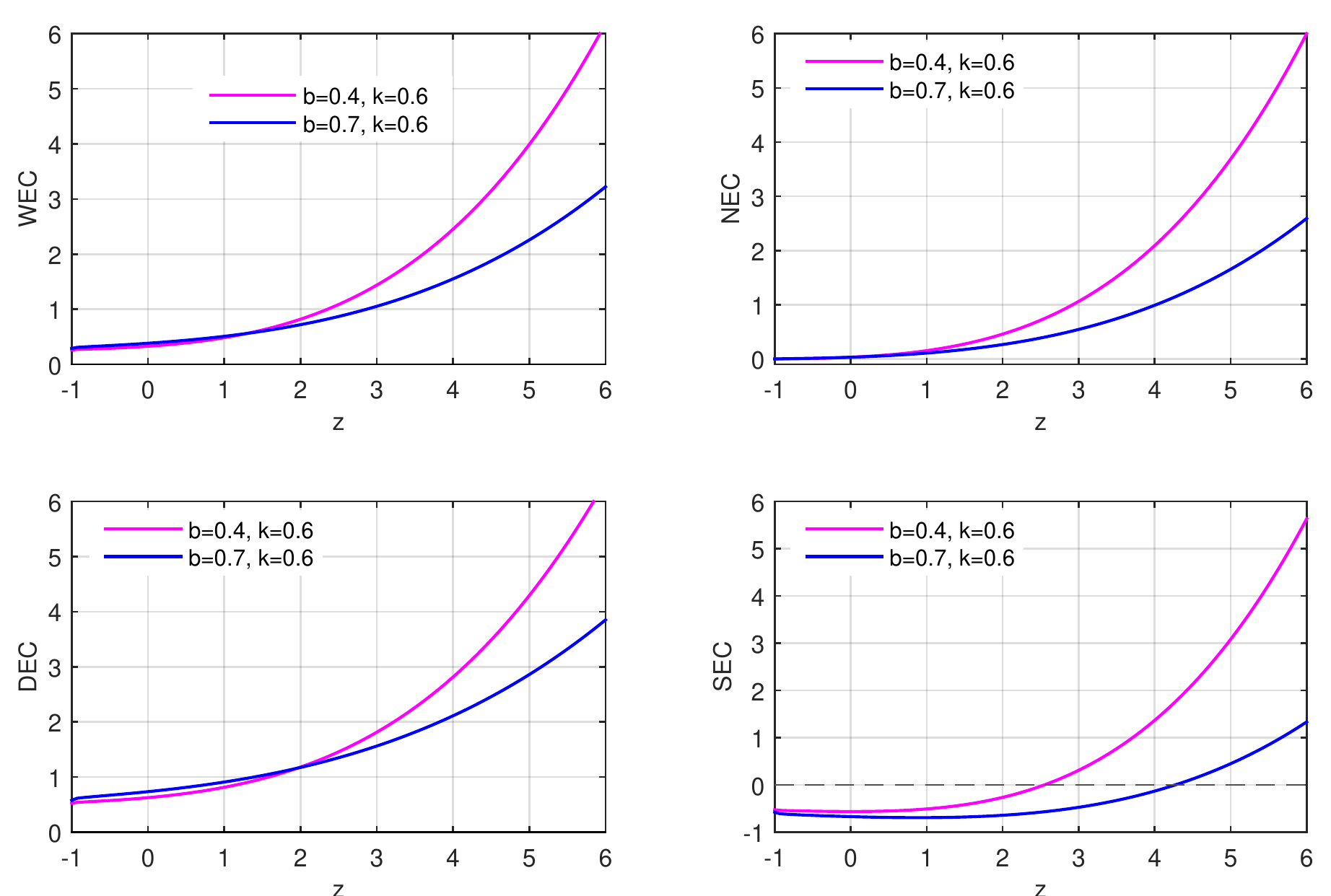}}
\caption{Evolution of energy conditions  in linear form of  $f (Q)$ gravity model versus redshift for the appropriate choice of constants $k=0.6$ and $b=0.4, 0.7$.}
\end{figure}
\subsection{Case-II: Non-linear case} The non-linear model takes the quadratic form i.e. $f(Q)=\alpha Q^{2}+\beta $ for $n=2$. Towards this model from equations (16) to (18), the expressions of physical parameters such as energy density, isotropic pressure and equation of state parameter along with energy conditions in terms of redshift as,\\
Isotropic pressure as,
\begin{equation}
p=\frac{\beta }{2}- 540 \alpha  \left[1+2\frac{\overset{.} {H}}{H^{2}} \right] H^{6 }, 
\end{equation}
Energy density as,
\begin{equation}
\rho =540 \alpha H^{6}- \frac{\beta }{2}, 
\end{equation}
\begin{figure}[H]
 \begin{minipage}{0.45\linewidth}
  \centerline{\includegraphics[scale=0.65]{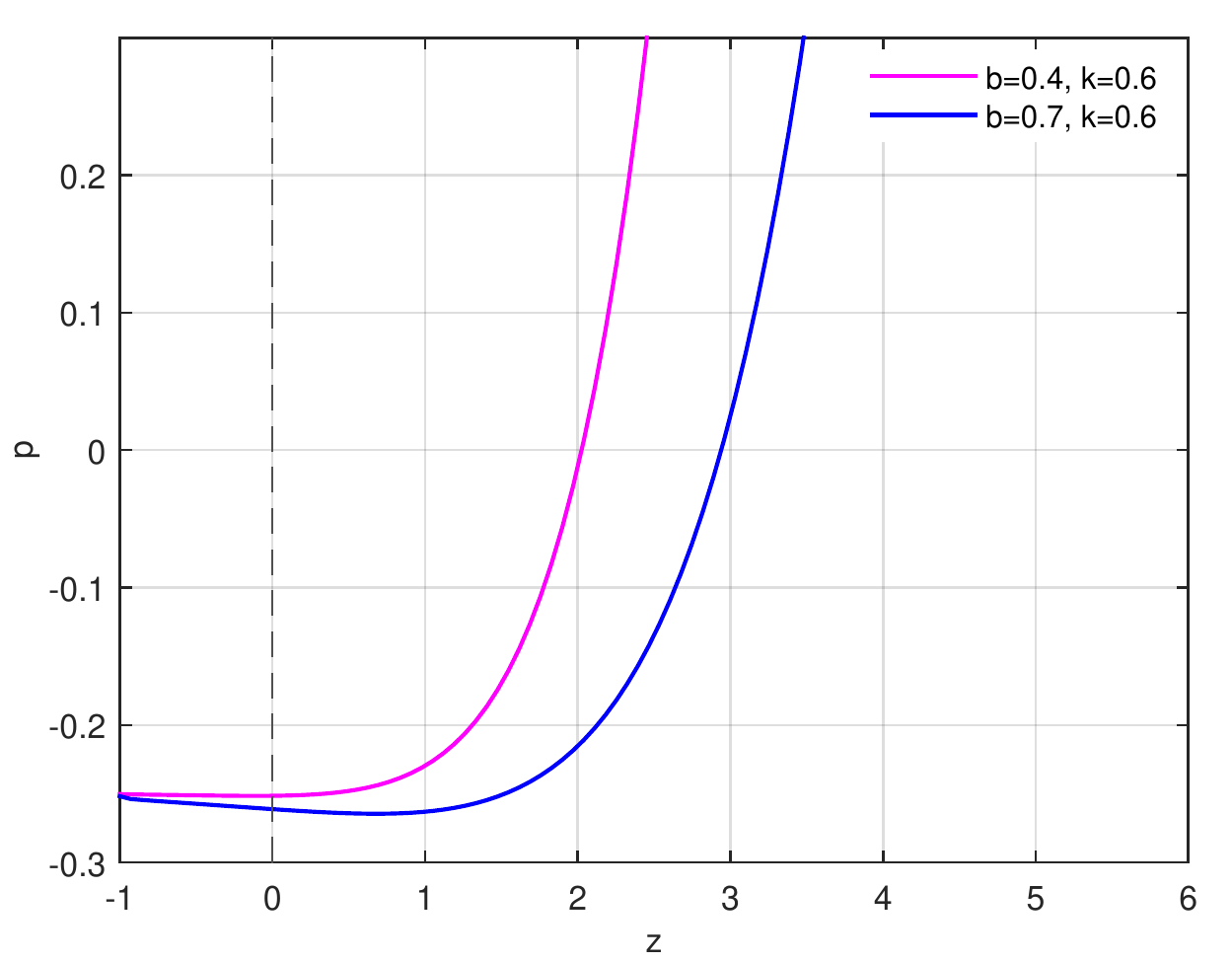}}
  \caption{Evolution of  isotropic pressure in non-linear form of $f (Q)$ gravity model versus redshift for the appropriate choice of constants $k=0.6$ and $b=0.4, 0.7$.}\label{fig3}
 \end{minipage}
\hfill
 \begin{minipage}{0.45\linewidth}
  \centerline{\includegraphics[scale=0.65]{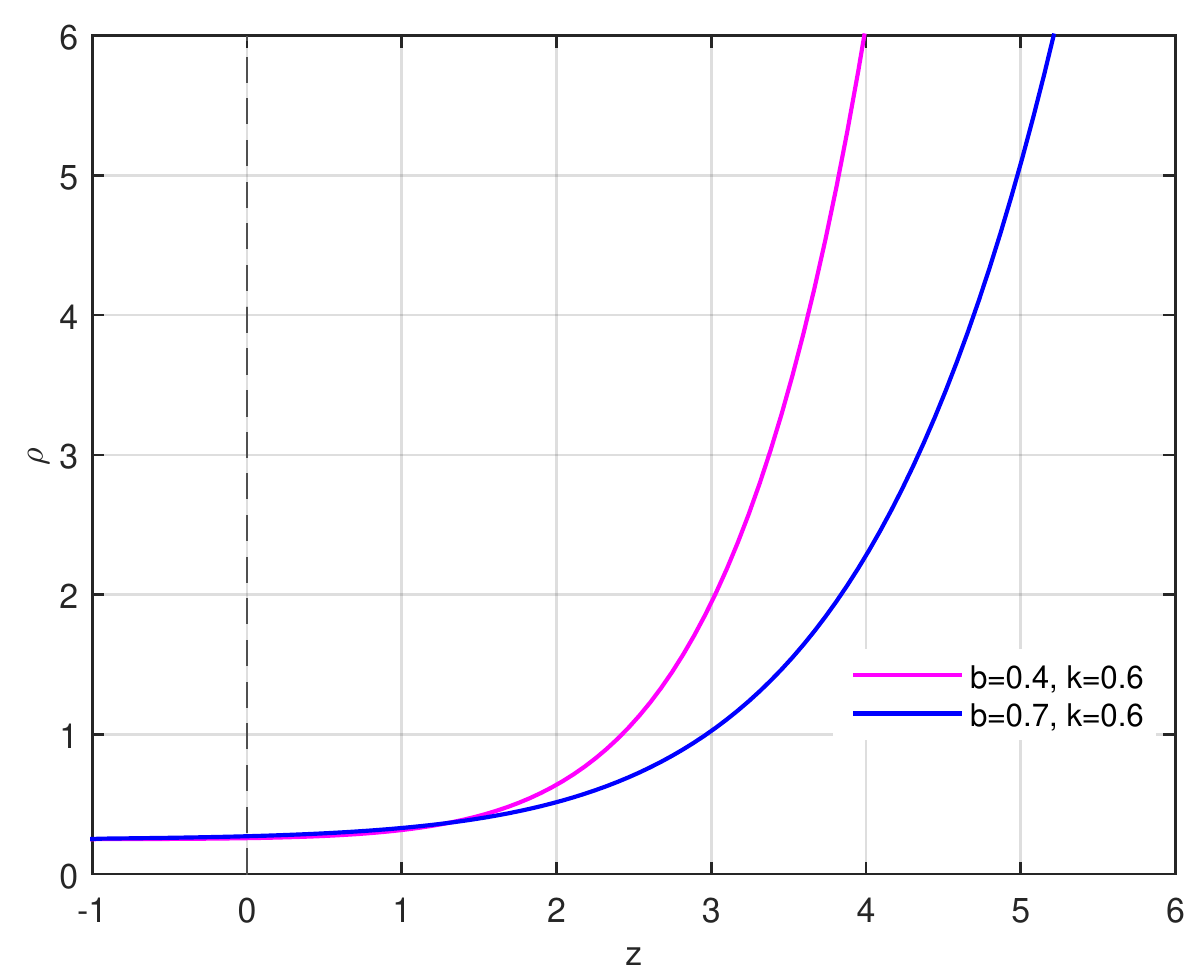}}
  \caption{Evolution of energy density in non-linear form of $f (Q)$ gravity model versus redshift for the appropriate choice of constants $k=0.6$ and $b=0.4, 0.7$.}\label{fig4}
 \end{minipage}
\end{figure}
The graphical illustration of the isotropic pressure and  the energy density in the derived non-linear model with respect to redshift is depicted in figs.(6) and (7) respectively by supposing the suitable choice of value of constants. From the figure it is observed that  the non-linear form of $f(Q)$ gravity model has the same behavior sa that of linear model in which an isotropic pressure has initially positive and with the evolution of the model it becomes negative, this negative pressure or repulsive force can be look upon as a source of inflation which is the exact capable tool to explain a cosmic acceleration also it is interesting to take note that the same behavior of isotropic pressure resumbles with the fact that the current Universe is an accelerating expansion due to large negative pressure (see fig. 6) while the energy density provides an empty universe for large time as shown in fig. (7) which confirms the fact that volume of the space increases because of the density of matter decreases as the Universe expands.\\
Equation of state parameter as,
\begin{equation}
\omega=\frac{2 \left(\beta- 540 \alpha H^{5} (H+2 \dot{H})\right)}{108 \alpha  H^{6}-\beta }
\end{equation}
The graphical performance of equation of state parameter of the derived non-linear model with respect to redshift is depicted in fig. (8) by supposing the suitable choice of values of constants.
\begin{figure}[H]
\centerline{\includegraphics[scale=0.9]{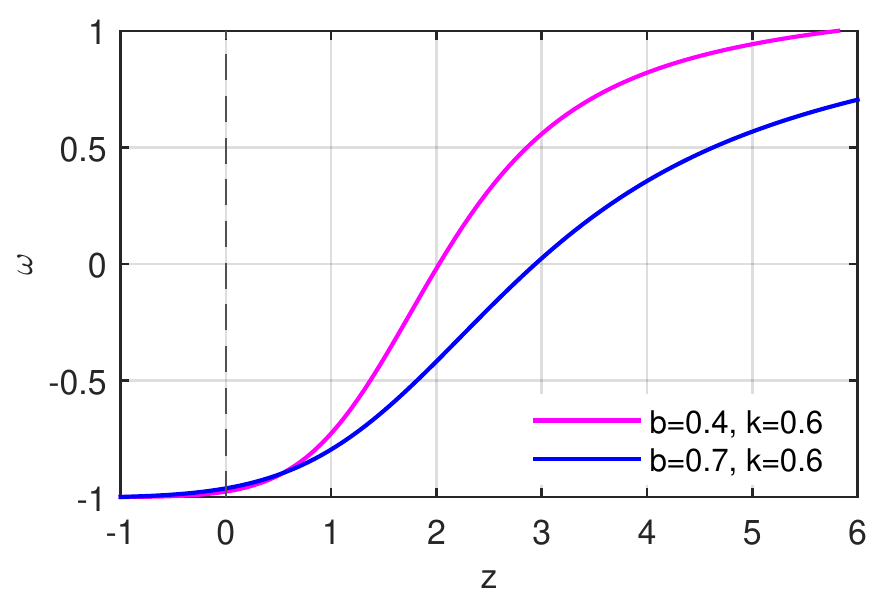}}
\caption{Evolution of equation of state parameter in non-linear form of $f (Q)$ gravity model versus redshift for the appropriate choice of constants $k=0.6$ and $b=0.4, 0.7$.}
\end{figure}
In a non-linear case, the behavior of $\omega$ is same as that the case of linear model which  converges to $\omega \rightarrow -1$ as time increases (see fig. 8). At present $z=0$ the model represents a quintessence region $\left( -1<\omega <-1/3\right) $ and  approaches towards the $\Lambda CDM$ model $\left( \omega =-1\right) $ in the future. This indicates that our model creates a very accelerated phase for very weak redshift values. Also, as the Planck and nine-year WMAP observational data expose that the value of equation of state parameter having range $\omega=-1.13_{-0.25}^{+0.24}$. In addition, we have observed that the present values of equation of state parameters for both the cases approximately accord with the Planck+WMAP observational data [33], which is an attractive result in this paper.\\
Energy conditions as,
\begin{equation}
WEC\Longleftrightarrow 540 \alpha  H^6-\frac{\beta }{2}, \left\{540 \alpha  H^6-540 \alpha H^6 \left(\frac{2 \dot{H}}{H^2}+1\right)\right\}
\end{equation}
\begin{equation}
NEC\Longleftrightarrow \left\{540 \alpha  H^6-540 \alpha H^6 \left(\frac{2 \dot{H}}{H^2}+1\right)\right\}
\end{equation}
\begin{equation}
DEC\Longleftrightarrow \left\{540 \alpha  H^6-\left| \frac{\beta }{2}-540 H^6 \left(\frac{2 \dot{H}}{H^2}+1\right) \alpha \right| -\frac{\beta }{2}\right\}
\end{equation}
\begin{equation}
SEC\Longleftrightarrow \left\{540 \alpha  H^6+3 \left(\frac{\beta }{2}-540 \alpha  H^6 \left(\frac{2 \dot{H}}{H^2}+1\right)\right)-\frac{\beta }{2}\right\}
\end{equation}
Evolution of energy conditions  in non-linear $f (Q)$ gravity model versus redshift for the appropriate choice of constants is presented in fig. (9). In fig. (9), we describe in detail the energy conditions  with $\alpha >0$ and $\beta <0$. Also, from this fig, we observed that $\rho \geq 0$, $\rho +p\geq 0$ and $\rho -p\geq 0$, which means that WEC, DEC and DEC are verified while SEC violates which leads to the acceleration of the Universe. Also, the SEC is violated in the present $\left( z=0\right) $ and the future $\left( z\rightarrow -1\right) $ in this model which predicts an accelerating expansin of the Universe.
\begin{figure}[H]
\centerline{\includegraphics[scale=0.9]{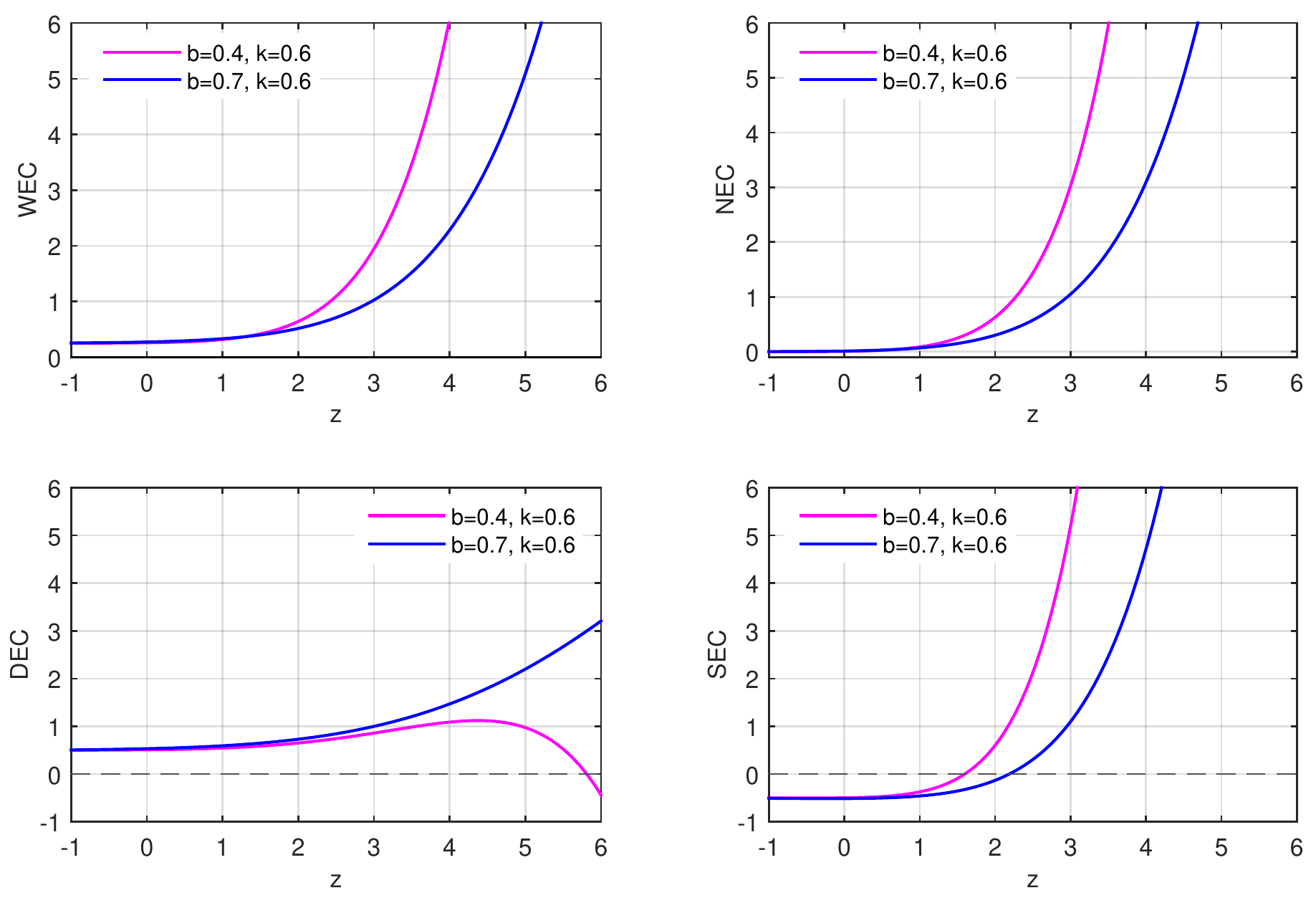}}
\caption{Evolution of energy conditions  in non-linear form of $f (Q)$ gravity model versus redshift for the appropriate choice of constants $k=0.6$ and $b=0.4, 0.7$.}
\end{figure}
\section{Jerk parameter}
At the beginning of the Universe, the density of dark energy was too low to overcome the gravity of the matter in the Universe, so the expansion was slow at the beginning. With the cosmic expansion, the density of matter decreases, and so the domination of matter will not lost and this is the era of dark energy domination that has already begun to faster cosmic expansion. Many researchers believe that the transition from the deceleration phase to an acceleration phase of the Universe is due to a cosmic jerk. This transition occurs in the universe for different models with a positive value of the jerk parameter and a negative value of the deceleration parameter. For example, the $\Lambda CDM$ model which is the most famous in cosmology has a constant jerk parameter $j=1$ [34]. The cosmic jerk parameter is a dimensionless quantity containing the third order derivative of the average scale factor with respect to the cosmic time and it is defined as $j=\frac{\overset{...}{a}}{aH^{3}}$. From equations (12) and (14), we get the expression for the jerk parameter as
\begin{equation}
j\left( z\right) =q+2q^{2}-\frac{\overset{.}{q}}{H}. 
\end{equation}
\begin{figure}[H]
\centerline{\includegraphics[scale=0.9]{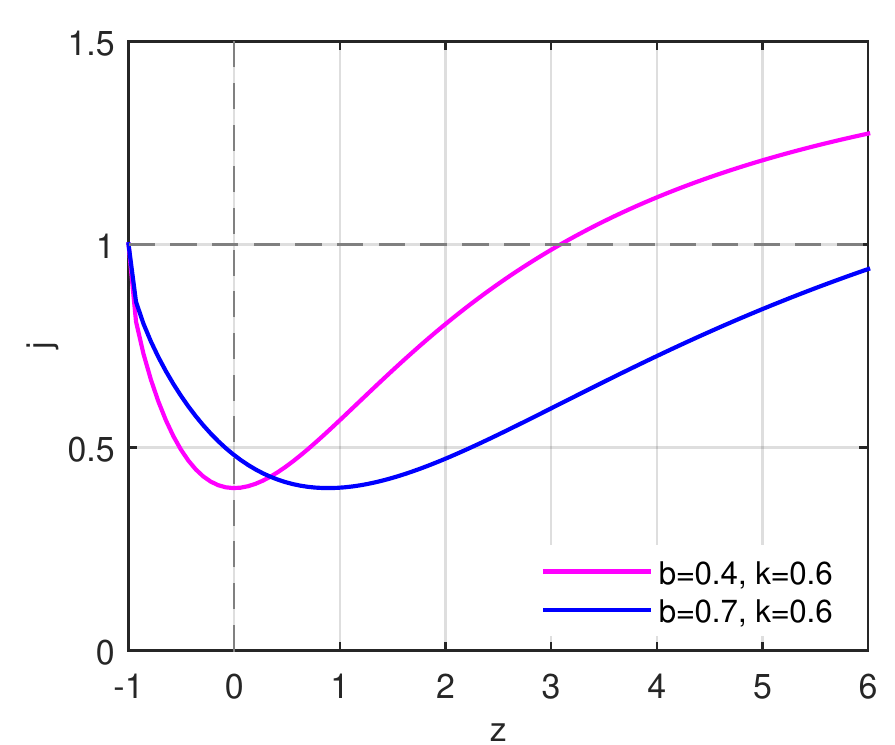}}
\caption{Evolution of Jerk parameter  in $f (Q)$ gravity model versus redshift for the appropriate choice of constants $k=0.6$ and $b=0.4, 0.7$.}
\end{figure}
Fig. (10) shows that the cosmic jerk parameter is positive throughout the entire life of the Universe and tends to $1$ at late times which validated that the model involve $\Lambda$CDM model with the strong justification of $\omega\rightarrow -1$.
\section{Statefinder parameters}
In order to understand the nature of dark energy and explain the accelerating expansion of the Universe, several models of dark energy have been developed. In order to distinguish between these models, an important parameters called Statefinder parameters were developed by Sahni et al. [35]. The statefinder parameters are related to the third order derivative of the mean scale factor. Also, the most important property of the statefinder pair is that $\left\{ r,s\right\}=\left\{ 1,0\right\} $ is a fixed point in the $\left\{ r,s\right\}$- plane of the spatially flat $\Lambda CDM$ model.
The statefinder parameters for our model are 
\begin{equation}
s=\frac{r-1}{3\left( q-\frac{1}{2}\right) },
\end{equation}
where $r=\frac{\overset{...}{a}}{aH^{3}}$.
\begin{figure}[H]
\centerline{\includegraphics[scale=0.9]{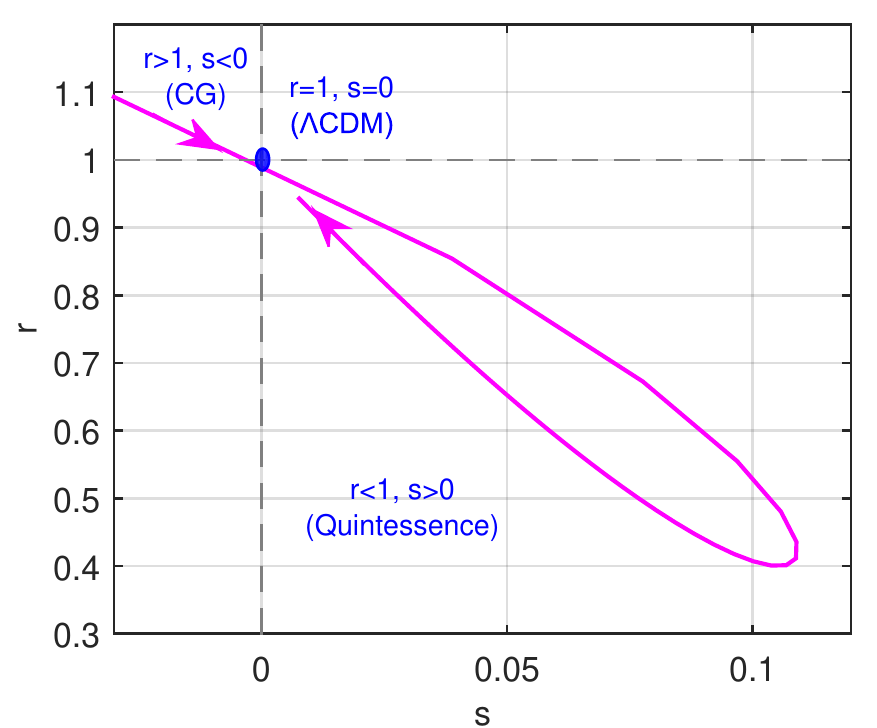}}
\caption{Evolution of statefinder parameters in $f (Q)$ gravity model versus redshift for the appropriate choice of constants $k=0.6$ and $b=0.4, 0.7$.}
\end{figure}
In fig. (11), we plot the evolution trajectories of our model at the $\left\{ r,s\right\}$- plane. We note that the evolution trajectory of our model ends at point $\left\{ r,s\right\} =\left\{ 1,0\right\} $. Also, we can see the behavior of $r$ and $s$ as $r\rightarrow 1$ and $s\rightarrow 0.$ So, our model corresponds to $\Lambda CDM$ model at late times.
\section{Conclusion}
In this work, we have studied the spatially homogeneous and isotropic cosmological model of FLRW in the framework of $f\left( Q\right) $ gravity model where $f\left( Q\right) =\alpha Q^{n}+\beta $. From the $f(Q)$ model here we analyzed two different cases like linear $\left( n=1\right) $ and quadratic $\left( n=2\right) $. We have obtained the exact solution of the field equations towards the hybrid expansion law together with the relation $t =\frac{kt_{0}}{b}f\left( z\right)$. In this analysis one observed the performance of the variation of the deceleration parameter from positive to negative which described that the Universe exhibits the transition from early deceleration phase to the current acceleration phase. Also, in these dynamical analysis the first major part is the analysis of linear and second is that the non-linear model of the said gravity. The outcomes in both the parts are as follows:\\
In both linear and non-linear models an isotropic pressure has initially positive and with the evolution of the model it becomes negative, this negative pressure or repulsive force can be look upon as a source of the iflation which is the exact capable tool to explain a cosmic acceleration also it is interesting to take note that the same behavior of isotropic pressure resumbles with the fact that the current Universe is an accelerating expansion due to large negative pressure while the energy density provides an empty Universe for large time which confirms the fact that volume of the space increases because of the density of matter decreases as the Universe expands. Also, in both the models the equation of state parameter is a function of redshift and converges to $\omega \rightarrow -1$ as time increases. The equation of state parameter represents a quintessence region $\left( -1<\omega <-1/3\right) $ at present $z=0$ and approache to the $\Lambda CDM$ model $\left( \omega =-1\right) $ in the future. This indicates that our model creates a very accelerated phase for very weak redshift values which is consistent with the recent observations. It is observed that in both models the WEC ($\rho \geq 0$) together with NEC ($\rho +p\geq 0$) and the DEC  ($\rho-\left\vert p\right\vert\ge 0$) are verified where as SEC ($\rho+3p\ge0$) violates. Therefore, the violation of SEC leads to the expansion of the Universe is accelerating. Also, it is seen that the SEC is violated at present $\left( z=0\right) $ and future $\left( z\rightarrow -1\right) $. Hence, as a result our model predicts an accelerating Universe.\\
Moreover, it is observed that the cosmic jerk parameter is positive throughout the entire evolution of the Universe and tends to $1$ at late times which confirmed the model involve $\Lambda$CDM mode with the strong justification of $\omega\rightarrow -1$ where as  statefinder parameters omits the value $r\rightarrow 1$ and $s \rightarrow 0$ which also confirmed that the model corresponds to $\Lambda CDM$ model at late times.

\acknowledgments
We are very much grateful to the honorary referee and the editor for the illuminating suggestions that have significantly improved our work in terms of research quality and presentation.\\

\textbf{Data availability} There are no new data associated with this article\\

\textbf{Declaration of competing interest} The authors declare that they have no known competing financial interests or personal relationships that could have appeared to influence the work reported in this paper.

\end{document}